\documentclass[journal,10pt]{IEEEtran}

\usepackage{cite}
\usepackage{graphicx}
\usepackage{epstopdf}
\usepackage{amsmath}
\usepackage{array}
\usepackage{mdwmath}
\usepackage{amssymb}
\usepackage{mdwtab}
\usepackage{stfloats}
\usepackage[tight,footnotesize]{subfigure}
\usepackage{amsmath,amsthm}
\usepackage{threeparttable}
\usepackage{color}
\usepackage{url}
\usepackage[ruled, linesnumbered]{algorithm2e}
\usepackage{multicol}
\usepackage{amssymb}
\usepackage{epstopdf}
\usepackage{bm}
\usepackage{flushend}

\begin{document}
	
\title{\LARGE Joint Optimization for Multi-User Transmissive RIS-MIMO Systems}

\author{Zhengwei Jiang, Yufeng Zhou, Xusheng Zhu, Wen Chen, Qingqing Wu, and Kai-Kit Wong \IEEEmembership{Fellow,~IEEE}
\thanks{
Z. Jiang is with China Unicom Shanghai Branch, Shanghai, China. (jiangzhenwei1@chinaunicom.cn.)}
\thanks{Y. Zhou, W. Chen and Q. Wu,  are with Department of Electronic Engineering, Shanghai Jiao Tong University, Shanghai, China. (ereaked@sjtu.edu.cn; wenchen@sjtu.edu.cn; qingqingwu@sjtu.edu.cn.)}
\thanks{X. Zhu and K. K. Wong are with Department of Electronic and Electrical Engineering, University College London, London, United Kingdom. (xusheng.zhu@ucl.ac.uk; kai-kit.wong@ucl.ac.uk.)}
}

\markboth{}
{}
\maketitle

\begin{abstract}
Transmissive reconfigurable intelligent surfaces (RIS) represent a transformative architecture for future wireless networks, enabling a paradigm shift from traditional costly base stations to low-cost, energy-efficient transmissive transmitters. This paper explores a downlink multi-user MIMO system where a transmissive RIS, illuminated by a single feed antenna, forms the core of the transmitter. The joint optimization of the RIS coefficient vector, power allocation, and receive beamforming in such a system is critical for performance but poses significant challenges due to the non-convex objective, coupled variables, and constant modulus constraints. To address these challenges, we propose a novel optimization framework. Our approach involves reformulating the sum-rate maximization problem into a tractable equivalent form and developing an efficient alternating optimization (AO) algorithm. This algorithm decomposes the problem into subproblems for the RIS coefficients, receive beamformers, and power allocation, each solved using advanced techniques including convex approximation and difference-of-convex programming. Simulation results demonstrate that our proposed method converges rapidly and achieves substantial sum-rate gains over conventional schemes, validating the effectiveness of our approach and highlighting the potential of transmissive RIS as a key technology for next-generation wireless systems.
\end{abstract}


\section{Introduction}

With the explosive development of the mobile Internet, there is an ever-increasing demand for higher data rates and greater transmission reliability in wireless communication \cite{zhu2025sssk}. While multiple-input multiple-output (MIMO) systems are a well-established solution for enhancing throughput by exploiting spatial channel resources, they suffer from a significant drawback: each antenna requires an expensive radio-frequency (RF) chain, leading to high hardware costs and power consumption, especially in massive MIMO deployments\cite{zhu2025discret}.

Reconfigurable intelligent surfaces (RIS) are emerging as a revolutionary technology for next-generation wireless networks\cite{tang2023trans}. Unlike traditional MIMO, an RIS is composed of numerous passive and low-cost elements that can manipulate electromagnetic (EM) waves via an intelligent controller \cite{zhu2024ond}. Each element can independently adjust the amplitude and/or phase of the incident signal, enabling the formation of highly focused beams towards a target \cite{zhu2024perdoubr}. These advantages have garnered significant research interest in wireless communications \cite{zhu2025risimp}.
Nevertheless, research into transmission-based RIS transceivers is still in its early stages, especially for the transmissive RIS transmitter \cite{liu2025lever,zhu2025trans}. Transmissive RIS-enabled transceivers overcome feed blockage and echo self-interference, enabling more efficient designs \cite{li2025trans,li2025toward,li2024towar} with lower power consumption and costs, making them highly promising for future communication networks. Although reflective RIS can also be used as a transceiver, transmissive RIS-based transceivers show greater promise and efficiency. Building upon prior work on transmissive RIS-based multi-antenna systems, we note that the transmissive RIS presents a potential solution to the high power and cost challenges of massive MIMO.

To address these challenges, we propose a novel joint optimization framework for a transmissive RIS-based multi-user MIMO system. Our approach reformulates the sum-rate maximization problem into a tractable equivalent form and develops an efficient alternating optimization (AO) algorithm. This algorithm iteratively solves three distinct subproblems for the RIS coefficient vector, receive beamforming matrix, and power allocation vector. We tackle the non-convexity of the RIS coefficient and receive beamforming subproblems by employing successive convex approximation (SCA), spherical coordinate parameterization, and techniques based on the difference-of-convex (DC) function property. The power allocation subproblem is formulated as a convex problem, which can be efficiently solved using standard optimization tools. Simulation results demonstrate that our proposed method not only converges rapidly but also achieves substantial sum-rate gains compared to conventional schemes.

\begin{figure}
  \centering
  \includegraphics[width=7cm]{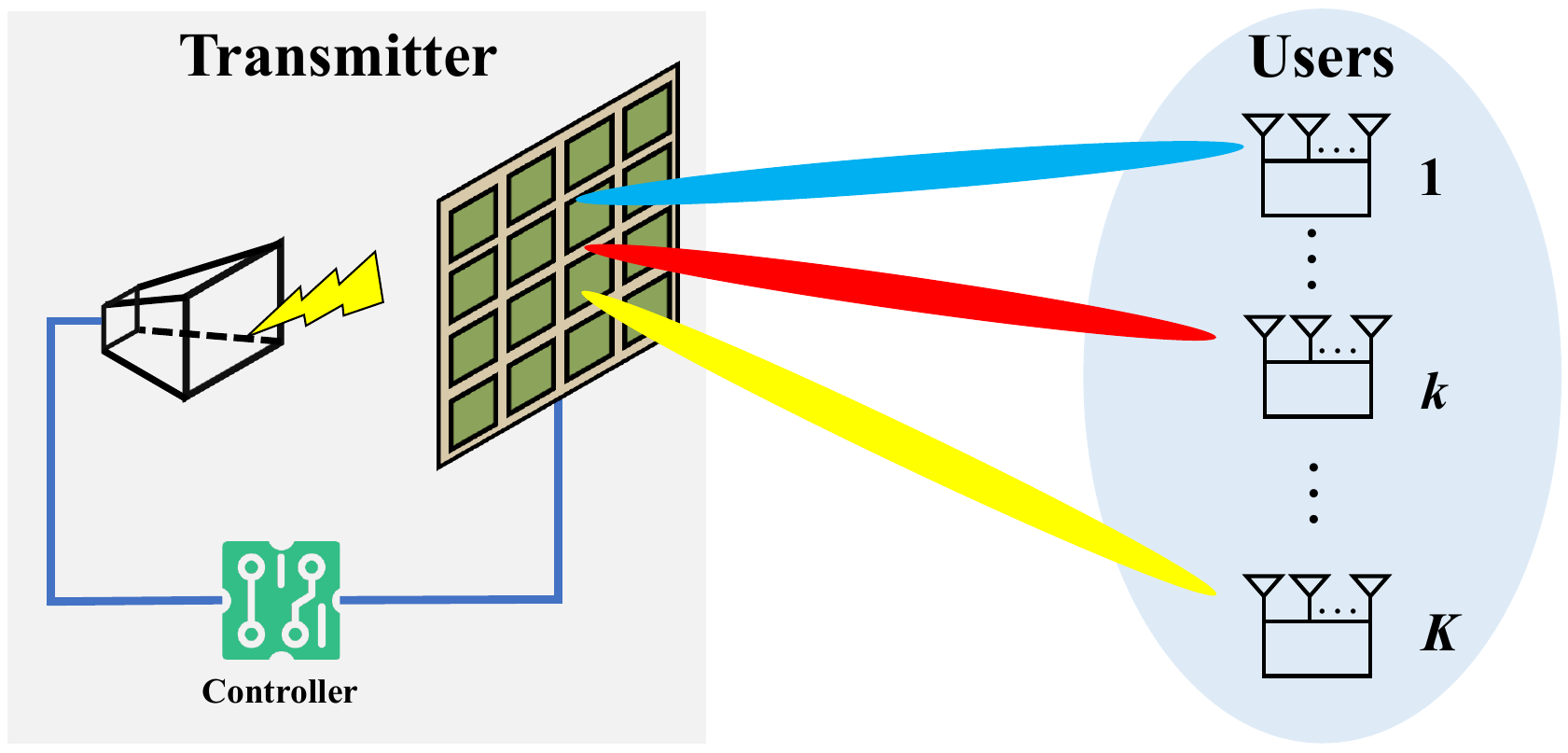}
  \caption{Transmissive RIS MIMO Networks.}\label{frame}
\end{figure}

\section{System Model}
As illustrated in Fig. \ref{frame}, we consider a downlink multi-user MIMO system that employs a transmissive RIS-based transmitter. This transmitter comprises a single feed antenna and a uniform linear array (ULA) of $M$ passive transmissive elements. The feed antenna continuously radiates EM waves towards the RIS, which dynamically controls the phase and amplitude of each transmitted signal via an intelligent controller. For analysis, we assume a total transmit power of $P_t$ and that the RIS transmits the incident signal with no reflection. The RIS coefficient vector is represented by $\mathbf{f}=\left[f_1,\cdots,f_M\right]^T\in \mathbb{C}^{M\times1}$, where each element $f_m = \beta_me^{j\theta_m}$ is defined by an amplitude $\beta_m\in[0, 1]$ and a phase shift $\theta_m \in [0, 2\pi)$. At the receiver side, there are $K$ users, each equipped with an $N$-antenna ULA.
\subsection{Channel Model}
\label{sec:channel_model}
The channel from the RIS to the $k$-th user is denoted as $\mathbf{H}_k \in \mathbb{C}^{N\times M}$ and is modeled as a Rician fading channel. This model captures both a dominant line-of-sight (LoS) component and a collection of non-line-of-sight (NLoS) paths. The channel matrix $\mathbf{H}_k$ can thus be expressed as
\begin{align}
	\mathbf{H}_k = \sqrt{\frac{C_0}{(d_k)^\alpha}}\left(\sqrt{\frac{\kappa}{\kappa+1}}\mathbf{H}_k^{LoS}+\sqrt{\frac{1}{\kappa+1}}\mathbf{H}_k^{NLoS}\right),\forall k,
\end{align}
where $C_0$ represents the channel gain at a reference distance of 1 meter, $d_k$ is the distance between the RIS and the $k$-th user, $\alpha$ is the path loss exponent, and $\kappa$ is the Rician factor.

The deterministic LoS component $\mathbf{H}_k^{LoS}$ is given by
\begin{align}
	\mathbf{H}_k^{LoS} = \mathbf{a}_N(\varphi_k)\mathbf{a}^H_M(\phi_k),\forall k,
\end{align}
where $\mathbf{a}_N(\varphi_k)$ and $\mathbf{a}_M(\phi_k)$ are the array response vectors at the user and the RIS, respectively. These vectors are defined as
\begin{align}
	&\mathbf{a}_N(\varphi_k) = [1, e^{j\frac{2\pi d}{\lambda}\sin(\varphi_k)}, \cdots, e^{j\frac{2\pi d}{\lambda}(N-1)\sin(\varphi_k)}]^T,\\
	&\mathbf{a}_M(\phi_k) = [1, e^{j\frac{2\pi d}{\lambda}\sin(\phi_k)}, \cdots, e^{j\frac{2\pi d}{\lambda}(M-1)\sin(\phi_k)}]^T,
\end{align}
where $d$ denotes the array element spacing, $\lambda$ is the wavelength, and $\varphi_k$ and $\phi_k$ are the angle of arrival (AoA) at the $k$-th user and the angle of departure (AoD) at the RIS, respectively.
The NLoS component $\mathbf{H}_k^{NLoS}$ follows a circularly symmetric complex Gaussian distribution, i.e., $\mathcal{CN}(0, \mathbf{I}_N)$. We assume a quasi-static flat fading channel, where $\mathbf{H}_k$ remains constant within each transmission slot.

\subsection{Signal Model}
The received signal at the $k$-th user is given by
\begin{align}
	\mathbf{y}_k = \mathbf{H}_k\mathbf{f}\mathbf{P}\mathbf{s}+\mathbf{n}_k,\forall k,
\end{align}
where $\mathbf{s}=[s_1, \dots, s_K]^T$ is the vector of transmitted signals, with $s_i \sim \mathcal{CN}(0, 1)$ for all $i$. The diagonal matrix $\mathbf{P}=\text{diag}(\sqrt{p_1},\dots,\sqrt{p_K})$ allocates transmit power $p_k$ to the $k$-th user. The additive white Gaussian noise (AWGN) at the $k$-th user is denoted by $\mathbf{n}_k$, which follows the distribution $\mathbf{n}_k \sim \mathcal{CN}(0, \sigma^2_k\mathbf{I}_N)$. The total transmit power is subject to the constraint $\sum_{k=1}^{K}p_k \leq P_t$.

The received signal can also be decomposed into the desired signal, inter-user interference, and noise
\begin{align}
	\mathbf{y}_k = \mathbf{H}_k\mathbf{f}\sqrt{p_k}s_k+\mathbf{H}_k\mathbf{f}\sum_{i\neq k}\sqrt{p_i}s_i+\mathbf{n}_k, \forall k.
\end{align}
After applying a normalized receiving beamforming vector $\mathbf{w}_k$ at the $k$-th user, the decoded signal $\hat{s}_k$ is
\begin{align}
	\hat{s}_k = \mathbf{w}_k^H\mathbf{H}_k\mathbf{f}\sqrt{p_k}s_k+\mathbf{w}_k^H\mathbf{H}_k\mathbf{f}\sum_{i\neq k}\sqrt{p_i}s_i+\mathbf{w}_k^H\mathbf{n}_k,\forall k.
\end{align}
Consequently, the signal-to-interference-plus-noise ratio (SINR) for the $k$-th user is given by
\begin{align}
	\text{SINR}_k = \frac{p_k\left|\mathbf{w}_k^H\mathbf{H}_k\mathbf{f}\right|^2}{\sum_{i\neq k}p_i\left|\mathbf{w}_k^H\mathbf{H}_k\mathbf{f}\right|^2+\sigma_k^2\left\|\mathbf{w}_k\right\|^2}, \forall k.
\end{align}
The achievable rate for the $k$-th user is then expressed as
\begin{align}
	R_k = \log_2(1+\text{SINR}_k),\forall k.
\end{align}
Finally, the achievable sum-rate of the proposed system is the sum of all user rates
\begin{align}
	R_{sum} = \sum_{k=1}^{K}R_k.
\end{align}

\subsection{Problem Formulation}
Our objective is to maximize the system's achievable sum-rate by jointly designing the transmissive RIS coefficient vector $\mathbf{f}$, the power allocation vector $\mathbf{p}$, and the receiving beamforming matrix $\mathbf{W} = \left[\mathbf{w}_1, \mathbf{w}_2, \dots, \mathbf{w}_K\right] \in \mathbb{C}^{N\times K}$. This optimization problem can be mathematically formulated as:
\begin{subequations}
	\begin{align}
		\label{P0}
		\text{(P0)}\quad \max_{\mathbf{f}, \mathbf{p}, \mathbf{W}} & \quad \sum_{k=1}^{K} \log_{2}\left(1 + {\rm SINR}_k\right),\\
		\text{s.t.} & \quad {\rm SINR}_k \geqslant \gamma_{th}, \quad k=1,\dots,K, \label{eq:qos}\\
		& \quad \left|f_m\right| \leqslant 1, \quad m=1,\dots,M, \label{eq:ris_constraint}\\
		& \quad p_k \geqslant 0, \quad k=1,\dots,K, \label{eq:power_nonnegative}\\
		& \quad \sum_{k=1}^{K} p_k \leqslant P_t, \label{eq:total_power}\\
		& \quad \Vert \mathbf{w}_k \Vert_2^2 = 1, \quad k=1,\dots,K. \label{eq:normalize_w}
	\end{align}
\end{subequations}
The constraints are defined as follows: Constraint \eqref{eq:qos} ensures that the quality-of-service (QoS) for each user meets a minimum SINR threshold, $\gamma_{th}$. Constraint \eqref{eq:ris_constraint} limits the magnitude of each RIS element's transmissive coefficient to a maximum value of 1. Constraints \eqref{eq:power_nonnegative} and \eqref{eq:total_power} represent the non-negativity and total transmit power budgets, respectively. Finally, constraint \eqref{eq:normalize_w} normalizes the receiving beamforming vector for each user.

The problem (P0) is a non-convex optimization problem due to the non-concave objective function and the highly coupled nature of the optimization variables. Consequently, it is challenging to find the global optimal solution directly. Therefore, we will develop an efficient algorithm to obtain a high-quality suboptimal solution.

\section{Proposed Solution}
\label{sec:proposed_solution}
To address the non-convex optimization problem (P0), we first transform it into an equivalent form (P1) by introducing auxiliary variables $\mathbf{r} = [r_1, \dots, r_K]$ and $\bm{\lambda} = [\lambda_1, \dots, \lambda_K]$, which represent the SINR-related terms and interference-plus-noise terms, respectively.
\subsection{Problem Transformation}
The equivalent problem (P1) is formulated as
\begin{subequations}
\label{eq:P1}
\begin{align}
\text{(P1)}\quad &\max_{\mathbf{f},\mathbf{p},\mathbf{W},\mathbf{r},\bm{\lambda}} \quad  \left( \prod_{k=1}^{K} r_k \right)^{\frac{1}{K}}, \\
\text{s.t.} \quad & (r_k - 1)\lambda_k \leq {p_k} \left| \mathbf{w}_k^H \mathbf{H}_k \mathbf{f} \right|^2/{\sigma^2}, \quad \forall k, \label{eq:P1b} \\
& \lambda_k \geq \sum\nolimits_{i \neq k} \frac{p_i}{\sigma^2} \left| \mathbf{w}_k^H \mathbf{H}_k \mathbf{f} \right|^2 + 1, \quad \forall k, \label{eq:P1c} \\
& r_k \geq 1 + \gamma_{\text{th}}, \quad \forall k, \label{eq:P1d} \\
& |f_m| \leq 1, \quad m = 1, \dots, M, \label{eq:P1e} \\
& p_k \geq 0, \quad \forall k, \label{eq:P1f} \\
& \sum\nolimits_{k=1}^{K} p_k \leq P_t, \label{eq:P1g} \\
& \| \mathbf{w}_k \|_2^2 = 1, \quad \forall k. \label{eq:P1h}
\end{align}
\end{subequations}
This transformation is based on the observation that for the optimal solution, the constraints \eqref{eq:P1b} and \eqref{eq:P1c} will hold with equality, effectively representing the SINR expression.

Due to the non-convexity of (P1) and the coupling among variables, we employ the AO framework to iteratively optimize three blocks of variables: the RIS coefficient vector $\mathbf{f}$, the receiving matrix $\mathbf{W}$, and the power allocation vector $\mathbf{p}$.

\begin{figure*}[t]
\begin{small}
\centering
\begin{equation}
\label{eq:gradient}
\nabla f_k(\bm{\theta}_k) =
\begin{pmatrix}
\displaystyle\sum_{n=1}^{2N} d^{(k)}_{n,n} 2\beta_{k,n} \frac{\partial \beta_{k,n}}{\partial \theta_{k,1}} + 2 \sum_{n=1}^{2N-1} \sum_{l=n+1}^{2N} d^{(k)}_{n,l} \left( \frac{\partial \beta_{k,n}}{\partial \theta_{k,1}} \beta_{k,l} + \beta_{k,n} \frac{\partial \beta_{k,l}}{\partial \theta_{k,1}} \right) \\
\vdots \\
\displaystyle\sum_{n=1}^{2N} d^{(k)}_{n,n} 2\beta_{k,n} \frac{\partial \beta_{k,n}}{\partial \theta_{k,2N-1}} + 2 \sum_{n=1}^{2N-1} \sum_{l=n+1}^{2N} d^{(k)}_{n,l} \left( \frac{\partial \beta_{k,n}}{\partial \theta_{k,2N-1}} \beta_{k,l} + \beta_{k,n} \frac{\partial \beta_{k,l}}{\partial \theta_{k,2N-1}} \right)
\end{pmatrix}
\end{equation}
\begin{equation}\label{eq:gradientx}
q_{i,j} \!=\! \sum_{n=1}^{2N} 2d^{(k)}_{n,n} \left( \frac{\partial^2 \beta_{k,n}}{\partial \theta_{k,i} \partial \theta_{k,j}} \!+ \!\frac{\partial \beta_{k,n}}{\partial \theta_{k,i}} \frac{\partial \beta_{k,n}}{\partial \theta_{k,j}} \right) \!+ \!\sum_{n=1}^{2N-1} \!\sum_{l=n+1}^{2N} 2d^{(k)}_{n,l} \left( \frac{\partial^2 \beta_{k,n}}{\partial \theta_{k,i} \partial \theta_{k,j}} \beta_{k,l} \!+\! \frac{\partial \beta_{k,n}}{\partial \theta_{k,i}} \frac{\partial \beta_{k,l}}{\partial \theta_{k,j}} \!+\! \frac{\partial \beta_{k,n}}{\partial \theta_{k,j}} \frac{\partial \beta_{k,l}}{\partial \theta_{k,i}} \!+ \!\beta_{k,n} \frac{\partial^2 \beta_{k,l}}{\partial \theta_{k,i} \partial \theta_{k,j}} \right).
\end{equation}
\end{small}
\hrule
\end{figure*}

\subsection{Optimizing RIS Coefficient Vector $\mathbf{f}$}
\label{subsec:optimize_f}
With $\mathbf{p}$ and $\mathbf{W}$ fixed, we optimize $\mathbf{f}$ and the auxiliary variables $\mathbf{r}, \bm{\lambda}$:
\begin{small}
\begin{subequations}
\label{eq:P20}
\begin{align}
\text{(P2.0)}\quad& \max_{\mathbf{f},\mathbf{r},\bm{\lambda}} \quad  \left( \prod_{k=1}^{K} r_k \right)^{\frac{1}{K}}, \\
\text{s.t.} \quad & r_k \lambda_k - \lambda_k \leq p_k \left| {\mathbf{w}_k^H \mathbf{H}_k}\mathbf{f}/{\sigma}  \right|^2, \quad \forall k, \label{eq:P2b} \\
& \lambda_k \geq \sum\nolimits_{i \neq k} p_i \left| {\mathbf{w}_k^H \mathbf{H}_k} \mathbf{f}/{\sigma} \right|^2 + 1, \quad \forall k, \label{eq:P2c} \\
& r_k \geq 1 + \gamma_{\text{th}}, \quad \forall k, \label{eq:P2d} \\
& |f_m| \leq 1, \quad m = 1, \dots, M. \label{eq:P2e}
\end{align}
\end{subequations}
\end{small}%
To handle the non-convex constraint \eqref{eq:P2b}, we introduce a real representation of $\mathbf{f}$
\begin{equation}
\bm{\alpha} = \begin{bmatrix} \Re(\mathbf{f}) \\ \Im(\mathbf{f}) \end{bmatrix},
\mathbf{A}_k = \begin{bmatrix}
\Re\left( \frac{\mathbf{w}_k^H \mathbf{H}_k}{\sigma} \right) & -\Im\left( \frac{\mathbf{w}_k^H \mathbf{H}_k}{\sigma} \right) \\
\Im\left( \frac{\mathbf{w}_k^H \mathbf{H}_k}{\sigma} \right) & \Re\left( \frac{\mathbf{w}_k^H \mathbf{H}_k}{\sigma} \right)
\end{bmatrix},
\end{equation}
so that
$
\left| \frac{\mathbf{w}_k^H \mathbf{H}_k}{\sigma} \mathbf{f} \right|^2 = \bm{\alpha}^T \mathbf{A}_k^T \mathbf{A}_k \bm{\alpha} \triangleq \bm{\alpha}^T \mathbf{B}_k \bm{\alpha}.
$
We then derive a concave lower bound for the right-hand side of \eqref{eq:P2b} using the first-order Taylor expansion
\begin{equation}
\begin{aligned}
\label{eq:lower_bound_f}
\bm{\alpha}^T \mathbf{B}_k \bm{\alpha} \geq &\bm{\alpha}^{(l)T} \mathbf{B}_k \bm{\alpha}^{(l)} + 2 \bm{\alpha}^{(l)T} \mathbf{B}_k (\bm{\alpha} - \bm{\alpha}^{(l)}) \\
&\triangleq g_k(\bm{\alpha}, \bm{\alpha}^{(l)}),
\end{aligned}
\end{equation}
where $\bm{\alpha}^{(l)}$ is the value from the previous iteration.
For the left-hand side of \eqref{eq:P2b}, we note that $r_k \lambda_k - \lambda_k$ is a difference-of-convex (DC) function. Using the inequality $xy \leq \frac{1}{4}(x+y)^2$, we obtain a convex upper bound as
\begin{equation}
\begin{aligned}
\label{eq:upper_bound_r_lambda}
r_k \lambda_k - \lambda_k \leq &\frac{1}{4}(r_k + \lambda_k)^2 - \frac{1}{4}[(r_k^{(l)} - \lambda_k^{(l)})^2 \\
&+ 2(r_k^{(l)} - \lambda_k^{(l)})(r_k - \lambda_k - r_k^{(l)} + \lambda_k^{(l)})] - \lambda_k \\
\triangleq& h(r_k, \lambda_k, r_k^{(l)}, \lambda_k^{(l)}),
\end{aligned}
\end{equation}
where $r_k^{(l)}$ and $\lambda_k^{(l)}$ are values from the previous iteration.
Thus, (P2.0) is approximated by the convex problem as
\begin{subequations}
\label{eq:P2}
\begin{align}
\text{(P2)}\quad &\max_{\bm{\alpha},\mathbf{r},\bm{\lambda}} \quad  \left( \prod_{k=1}^{K} r_k \right)^{\frac{1}{K}}, \\
\text{s.t.} \quad & h(r_k, \lambda_k, r_k^{(l)}, \lambda_k^{(l)}) \leq p_k g_k(\bm{\alpha}, \bm{\alpha}^{(l)}), \quad \forall k, \\
& \sum\nolimits_{i \neq k} p_i \bm{\alpha}^T \mathbf{B}_k \bm{\alpha} + 1 - \lambda_k \leq 0, \quad \forall k, \\
& r_k \geq 1 + \gamma_{\text{th}}, \quad \forall k, \\
& \left\| [\alpha_m, \alpha_{M+m}]^T \right\|_2^2 \leq 1, \quad m = 1, \dots, M.
\end{align}
\end{subequations}
The constraint $\left\| [\alpha_m, \alpha_{M+m}]^T \right\|_2^2 \leq 1$ ensures that $|f_m| \leq 1$ for each element.

\subsection{Optimizing Receiving Matrix $\mathbf{W}$}
\label{subsec:optimize_W}
With $\mathbf{f}$ and $\mathbf{p}$ fixed, we optimize $\mathbf{W}$ and the auxiliary variables as
\begin{subequations}
\label{eq:P30}
\begin{align}
\text{(P3.0)}\quad &\max_{\mathbf{W},\mathbf{r},\bm{\lambda}} \quad  \left( \prod_{k=1}^{K} r_k \right)^{\frac{1}{K}}, \\
\text{s.t.} \quad & r_k \lambda_k - \lambda_k \leq p_k \left| \mathbf{w}_k^H \frac{\mathbf{H}_k \mathbf{f}}{\sigma} \right|^2, \quad \forall k, \label{eq:P30b} \\
& \lambda_k \geq \sum_{i \neq k} p_i \left| \mathbf{w}_k^H \frac{\mathbf{H}_k \mathbf{f}}{\sigma} \right|^2 + 1, \quad \forall k, \label{eq:P30c} \\
& r_k \geq 1 + \gamma_{\text{th}}, \quad \forall k, \label{eq:P30d} \\
& \| \mathbf{w}_k \|_2^2 = 1, \quad \forall k. \label{eq:P30e}
\end{align}
\end{subequations}
We define a real representation for $\mathbf{w}_k$ as
\begin{equation}
\bm{\beta}_k = \begin{bmatrix} \Re(\mathbf{w}_k) \\ \Im(\mathbf{w}_k) \end{bmatrix},
\mathbf{C}_k = \begin{bmatrix}
\Re\left( \frac{\mathbf{f}^H \mathbf{H}_k^H}{\sigma} \right) & -\Im\left( \frac{\mathbf{f}^H \mathbf{H}_k^H}{\sigma} \right) \\
\Im\left( \frac{\mathbf{f}^H \mathbf{H}_k^H}{\sigma} \right) & \Re\left( \frac{\mathbf{f}^H \mathbf{H}_k^H}{\sigma} \right)
\end{bmatrix},
\end{equation}
so that
$
\left| \mathbf{w}_k^H \frac{\mathbf{H}_k \mathbf{f}}{\sigma} \right|^2 = \bm{\beta}_k^T \mathbf{C}_k^T \mathbf{C}_k \bm{\beta}_k \triangleq \bm{\beta}_k^T \mathbf{D}_k \bm{\beta}_k.
$
To handle the unit-norm constraint \eqref{eq:P30e}, we parameterize $\bm{\beta}_k$ using spherical coordinates
$
\bm{\theta}_k = [\theta_{k,1}, \dots, \theta_{k,2N-1}]^T, \quad \theta_{k,i} \in [-\pi/2, \pi/2], \; \theta_{k,2N-1} \in [-\pi, \pi],
$
and define
\begin{equation}
\bm{\beta}_k =
\begin{bmatrix}
\sin(\theta_{k,1}) \\
\cos(\theta_{k,1}) \sin(\theta_{k,2}) \\
\vdots \\
\cos(\theta_{k,1}) \cdots \cos(\theta_{k,2N-2}) \sin(\theta_{k,2N-1}) \\
\cos(\theta_{k,1}) \cdots \cos(\theta_{k,2N-2}) \cos(\theta_{k,2N-1})
\end{bmatrix}.
\end{equation}
This parameterization automatically satisfies $\| \bm{\beta}_k \|_2 = 1$.
Let
$
f_k(\bm{\theta}_k) = \bm{\beta}_k^T \mathbf{D}_k \bm{\beta}_k,
$
we derive its gradient $\nabla f_k(\bm{\theta}_k)$ in \eqref{eq:gradient} and Hessian $\nabla^2 f_k(\bm{\theta}_k)$ as $\nabla^2 f_k(\bm{\theta}_k) = (q_{i,j})_{(2N-1) \times (2N-1)}$, where $q_{i,j}$ is in \eqref{eq:gradientx}.

Using Taylor expansion, we construct convex approximations as
\begin{small}
\begin{equation}
\setcounter{equation}{23}
\begin{aligned}
\label{eq:m1}
m_k^1(\bm{\theta}_k, \bm{\theta}_k^{(l)}) = &f_k(\bm{\theta}_k^{(l)}) + \nabla f_k(\bm{\theta}_k^{(l)})^T (\bm{\theta}_k - \bm{\theta}_k^{(l)}) \\
&+ \frac{1}{2} \delta_k^1 \| \bm{\theta}_k - \bm{\theta}_k^{(l)} \|_2^2,
\end{aligned}
\end{equation}
\end{small}
\begin{small}
\begin{equation}
\begin{aligned}
\label{eq:m2}
m_k^2(\bm{\theta}_k, \bm{\theta}_k^{(l)}) = &-f_k(\bm{\theta}_k^{(l)}) - \nabla f_k(\bm{\theta}_k^{(l)})^T (\bm{\theta}_k - \bm{\theta}_k^{(l)}) \\
&+ \frac{1}{2} \delta_k^2 \| \bm{\theta}_k - \bm{\theta}_k^{(l)} \|_2^2,
\end{aligned}
\end{equation}
\end{small}
where $\delta_k^1, \delta_k^2$ are chosen such that $\delta_k^1 \mathbf{I} \succeq \nabla^2 f_k(\bm{\theta}_k)$ and $\delta_k^2 \mathbf{I} \succeq -\nabla^2 f_k(\bm{\theta}_k)$ for all $\bm{\theta}_k$ in the domain. A sufficient condition is to set
\begin{equation}
\delta_k^1 = \delta_k^2 = \left\| 4 \sum_{n=1}^{2N} d^{(k)}_{n,n} + 8 \sum_{n=1}^{2N-1} \sum_{l=n+1}^{2N} d^{(k)}_{n,l} \right\| (2N - 1),
\end{equation}
which upper-bounds the Frobenius norm of the Hessian.
As such, the resulting convex problem is
\begin{subequations}
\label{eq:P3}
\begin{align}
\text{(P3)}\quad &\max_{\bm{\theta},\mathbf{r},\bm{\lambda}} \quad  \left( \prod_{k=1}^{K} r_k \right)^{\frac{1}{K}}, \\
\text{s.t.} \quad & h(r_k, \lambda_k, r_k^{(l)}, \lambda_k^{(l)}) + p_k m_k^2(\bm{\theta}_k, \bm{\theta}_k^{(l)}) \leq 0,  \forall k, \\
& \sum_{i \neq k} p_i m_k^1(\bm{\theta}_k, \bm{\theta}_k^{(l)}) + 1 - \lambda_k \leq 0, \forall k, \\
& r_k \geq 1 + \gamma_{\text{th}}, \forall k.
\end{align}
\end{subequations}

\subsection{Optimizing Power Allocation $\mathbf{p}$}
\label{subsec:optimize_p}
With $\mathbf{f}$ and $\mathbf{W}$ fixed, we optimize $\mathbf{p}$ as
\begin{small}
\begin{subequations}
\label{eq:P4}
\begin{align}
\text{(P4)}\quad & \max_{\mathbf{p},\mathbf{r},\bm{\lambda}} \quad  \left( \prod_{k=1}^{K} r_k \right)^{\frac{1}{K}}, \\
\text{s.t.} \quad & h(r_k, \lambda_k, r_k^{(l)}, \lambda_k^{(l)}) \leq \left| \frac{\mathbf{w}_k^H \mathbf{H}_k \mathbf{f}}{\sigma} \right|^2 p_k, \quad \forall k, \\
& \lambda_k \geq \sum_{i \neq k} p_i \left| \frac{\mathbf{w}_k^H \mathbf{H}_k \mathbf{f}}{\sigma} \right|^2 + 1, \quad \forall k, \\
& r_k \geq 1 + \gamma_{\text{th}}, \quad \forall k, \\
& p_k \geq 0, \quad \forall k, \\
& \sum_{k=1}^{K} p_k \leq P_t.
\end{align}
\end{subequations}
\end{small}%
It is obvious that (P4) a convex problem since all constraints are linear in $\mathbf{p}$ and the auxiliary variables, and can be efficiently solved using standard convex optimization tools.

\begin{algorithm}[t]
\caption{Proposed Transmissive RIS MIMO Optimization Algorithm}
\label{alg:main}
\textbf{Initialize:} $\bm{\alpha}^{(0)}$, $\bm{\theta}^{(0)}$, $\mathbf{p}^{(0)}$; choose $\epsilon$, $\epsilon_1$, $\epsilon_2$; set max iters $\vartheta$, $\vartheta_1$, $\vartheta_2$; $n=0$, $l_1=0$, $l_2=0$.\\

Calculate $\mathbf{W}^{(n)}$ and $R_{sum}^{(n)}$.\\

\While{$n \leq \vartheta$ and $|R_{sum}^{(n)} - R_{sum}^{(n-1)}| > \epsilon$}{
	Set $\hat{\bm{\alpha}}^{(0)} \gets \bm{\alpha}^{(n)}$, $l_1 = 0$.\\
	\While{$l_1 \leq \vartheta_1$ and $\Vert \hat{\bm{\alpha}}^{(l_1)} - \hat{\bm{\alpha}}^{(l_1-1)} \Vert_2 > \epsilon_1$}{
		$l_1 \gets l_1 + 1$.\\
		Solve (P2) with $\mathbf{W}^{(n)}$, $\mathbf{p}^{(n)}$, $\hat{\bm{\alpha}}^{(l_1-1)}$ to get $\hat{\bm{\alpha}}^{(l_1)}$.
	}	
	Set $\bm{\alpha}^{(n+1)} \gets \hat{\bm{\alpha}}^{(l_1)}$ and compute $\mathbf{f}^{(n+1)}$.\\

	Set $\hat{\bm{\theta}}^{(0)} \gets \bm{\theta}^{(n)}$, $l_2 = 0$.\\
	\While{$l_2 \leq \vartheta_2$ and $\Vert \hat{\bm{\theta}}^{(l_2)} - \hat{\bm{\theta}}^{(l_2-1)} \Vert_2 > \epsilon_2$}{
		$l_2 \gets l_2 + 1$.\\
		Solve (P3) with $\mathbf{f}^{(n+1)}$, $\mathbf{p}^{(n)}$, $\hat{\bm{\theta}}^{(l_2-1)}$ to get $\hat{\bm{\theta}}^{(l_2)}$.
	}	
	Set $\bm{\theta}^{(n+1)} \gets \hat{\bm{\theta}}^{(l_2)}$ and compute $\mathbf{W}^{(n+1)}$.\\

	Solve (P4) with $\mathbf{f}^{(n+1)}$, $\mathbf{W}^{(n+1)}$ to get $\mathbf{p}^{(n+1)}$.\\

	Calculate $R_{sum}^{(n+1)}$ and set $n \gets n + 1$.
}
\end{algorithm}

\subsection{Algorithm Complexity}
\label{subsec:complexity}

The computational complexity of Algorithm \ref{alg:main} is dominated by the three convex subproblems:
1) (P2): $\mathcal{O}\left( [2(M+K)]^{3.5} \ln(1/\varepsilon) \right)$
2) (P3): $\mathcal{O}\left( [2(N+1)K]^{3.5} \ln(1/\varepsilon) \right)$
3) (P4): $\mathcal{O}\left( (3K)^{3.5} \ln(1/\varepsilon) \right)$.
Let $l$, $l_1$, $l_2$ be the iteration numbers for the outer loop and inner loops for (P2) and (P3), respectively. Total complexity is $\mathcal{O}\left( l \ln(1/\varepsilon) \left[ l_1 [2(M+K)]^{3.5} + l_2 [2(N+1)K]^{3.5} + (3K)^{3.5} \right] \right)$.

\begin{figure*}[ht]
\centering
\begin{minipage}[t]{0.32\textwidth}
\centering
\includegraphics[width=\linewidth]{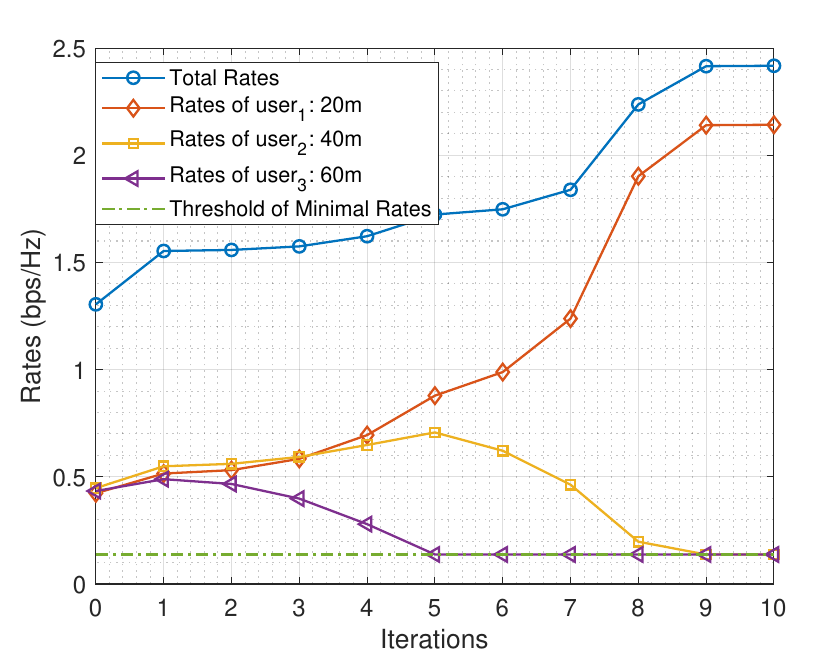}
\caption{Convergence performance.}
\label{fig:convergence}
\end{minipage}
\hfill
\begin{minipage}[t]{0.32\textwidth}
\centering
\includegraphics[width=\linewidth]{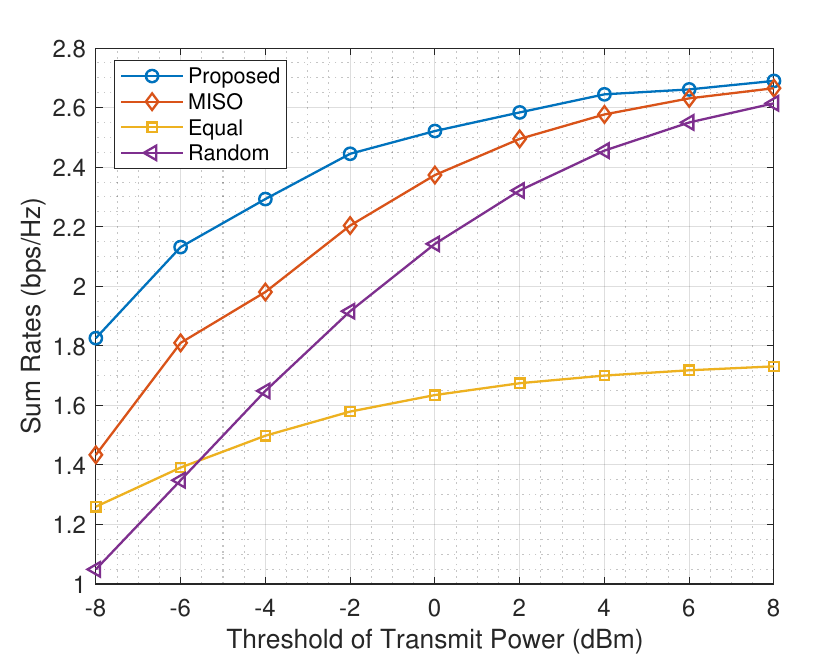}
\caption{Sum rates versus transmit power threshold.}
\label{fig:power}
\end{minipage}
\hfill
\begin{minipage}[t]{0.32\textwidth}
\centering
\includegraphics[width=\linewidth]{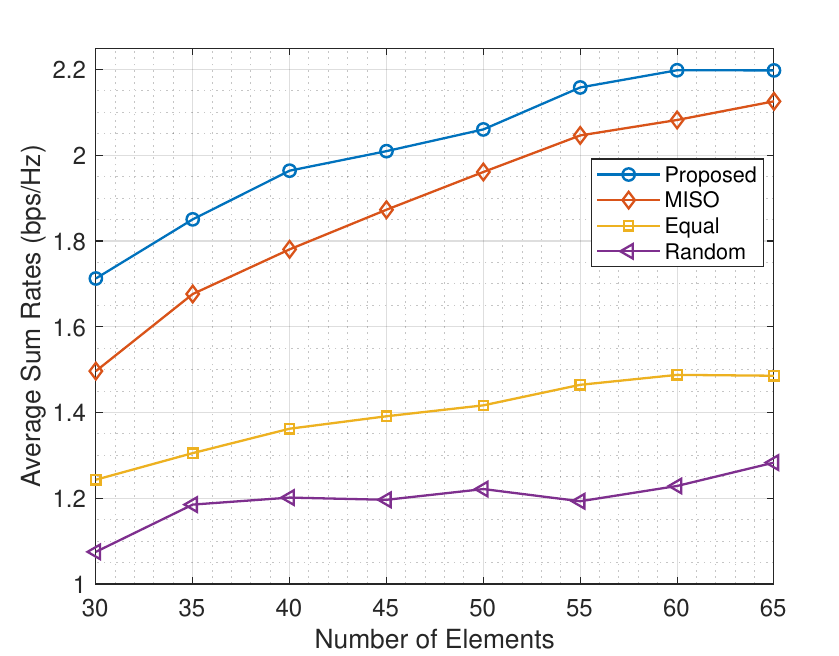}
\caption{Average sum rates versus RIS elements.}
  \label{fig:elements}
\end{minipage}
\end{figure*}

\section{Simulation Results}
\label{sec:simulation}
In this section, a transmissive RIS with \(M = 30\) serves \(K = 3\) users, each with \(N = 4\) receive antennas. The users are uniformly distributed in a ring-shaped area around the RIS, with distances \(d_k \sim \mathcal{U}[20, 60]\) meters. The channel follows Rician fading (Section~\ref{sec:channel_model}) with parameters: reference channel gain \(C_0 = -30\) dB, Rician factor \(\kappa = 3\), and path-loss exponent \(\alpha = 2.2\). The transmit power is \(P_t = -8\) dBm, the noise power at each user is \(\sigma_k^2 = -50\) dBm, and the minimum SINR requirement is \(\gamma_{\text{th}} = 0.1\).
We compare the proposed algorithm with the following baselines:
1) \textbf{MISO}: Each user has a single antenna.
2) \textbf{Equal power}: Power is equally allocated among users.
3) \textbf{Random phase}: The RIS phase shifts are randomly chosen from \([0, 2\pi)\).

Fig.~2 illustrates the convergence performance of the proposed algorithm. It is observed that the sum-rate increases monotonically and converges within 9 iterations, demonstrating the efficiency and stability of the proposed alternating optimization framework. Notably, while User 1's rate increases consistently, the rates of Users 2 and 3 initially increase and then decrease until their minimum SINR constraints are satisfied. This behavior indicates that the algorithm allocates resources preferentially to the user with the best channel condition while still guaranteeing the QoS requirements for all users.

Fig.~3 shows the sum-rate versus the maximum transmit power \(P_t\). As expected, the sum-rate increases with \(P_t\) for all schemes. The proposed algorithm consistently outperforms all baselines across the entire power range. At low transmit power, the random-phase scheme performs worse than the equal-power scheme due to the lack of optimized phase alignment. However, as \(P_t\) increases, the random-phase scheme eventually surpasses the equal-power scheme, indicating that inter-user interference becomes the dominant limiting factor when power is sufficient. The MISO scheme performs worse than the MIMO case due to reduced receive diversity. The proposed scheme achieves the highest sum-rate by jointly optimizing the RIS coefficients, power allocation, and receive beamforming.

Fig.~4 depicts the sum-rate versus the number of RIS elements \(M\). The sum-rate improves with \(M\) for all schemes, as a larger RIS provides higher beamforming gain and greater spatial diversity. The random-phase scheme exhibits the slowest growth due to the lack of phase optimization. The proposed algorithm achieves the highest sum-rate for all values of $M$, demonstrating its ability to fully exploit the increased degrees of freedom offered by additional RIS elements.

\section{Conclusions}
This paper addresses the key challenges in transmissive RIS-based MIMO systems, where the joint optimization of system parameters presents significant difficulties due to non-convex objectives and practical constraints. We develop an efficient optimization framework that transforms the original problem and employs an alternating optimization approach, systematically solving subproblems for RIS coefficients, receive beamforming, and power allocation through convex approximation techniques. Simulation results demonstrate substantial performance gains over conventional schemes, validating our approach and establishing transmissive RIS as a promising architecture for future wireless systems. This work provides important insights into the co-design methodology for RIS-based transmitters and lays the foundation for further research.

\end{document}